\documentclass[aps, prx, notitlepage, citeautoscript, superscriptaddress, reprint]{revtex4-2}

\usepackage{xr}
\usepackage{amsmath}
\usepackage{amsfonts,amssymb,amsthm,amsxtra,physics,dsfont}
\usepackage[]{graphicx}
\usepackage{grffile}
\usepackage{mathrsfs}
\usepackage{framed}
\usepackage{bbm}
\usepackage{bm}
\usepackage{braket}
\usepackage{xcolor}
\usepackage{mathtools}
\usepackage{tikz}
\usetikzlibrary{quantikz}
\usepackage[bookmarks=false,colorlinks=true,urlcolor=blue,citecolor=blue,linkcolor=blue]{hyperref}
\usepackage{xcolor, soul}
\sethlcolor{yellow}
\usepackage{private}

\usepackage{makecell}
\usepackage{lipsum}
\usepackage{pgfplots}
\usepackage{comment}

\usepackage[capitalize]{cleveref}
\usepackage{algorithm}
\usepackage{algpseudocode}

\preprint{APS/123-QED}

\pgfplotsset{compat=1.18}

\begin{document}
\title{Powerful Quantum Circuit Resizing with Resource Efficient Synthesis} 


\author{Siyuan Niu}
\email{siyuanniu@lbl.gov}
\affiliation{Computational Research Division, Lawrence Berkeley National Laboratory, Berkeley, California 94720, USA}

\author{Akel Hashim}
\affiliation{Quantum Nanoelectronics Laboratory, Department of Physics,
University of California, Berkeley, California 94720, USA}
\affiliation{Computational Research Division, Lawrence Berkeley National Laboratory, Berkeley, California 94720, USA}

\author{Costin Iancu}
\affiliation{Computational Research Division, Lawrence Berkeley National Laboratory, Berkeley, California 94720, USA}

\author{Wibe Albert de Jong}
\affiliation{Computational Research Division, Lawrence Berkeley National Laboratory, Berkeley, California 94720, USA}

\author{Ed Younis}
\email{edyounis@lbl.gov}
\affiliation{Computational Research Division, Lawrence Berkeley National Laboratory, Berkeley, California 94720, USA}

\begin{abstract}
\label{sec:abstract}
In the noisy intermediate-scale quantum era, mid-circuit measurement and reset operations facilitate novel circuit optimization strategies by reducing a circuit's qubit count in a method called resizing. This paper introduces two such algorithms. The first one leverages gate-dependency rules to reduce qubit count by 61.6\% or 45.3\% when optimizing depth as well. Based on numerical instantiation and synthesis, the second algorithm finds resizing opportunities in previously unresizable circuits via dependency rules and other state-of-the-art tools. This resizing algorithm reduces qubit count by 20.7\% on average for these previously impossible-to-resize circuits.

\end{abstract}
    
\maketitle

\section{Introduction}
\label{sec:introduction}

Quantum computing promises to address classically intractable
problems, particularly in chemistry, optimization, machine learning, and
physical simulations. With more fields of application continuously being
discovered and quantum hardware consistently improving, the future of
quantum computers is auspicious. Nevertheless, today's machines are
still considered Noisy Intermediate-Scale Quantum (NISQ) devices because
they consist of a few hundred imperfect qubits. There is potential for executing quantum algorithms on NISQ machines to outperform their classical counterparts, however, only if we
utilize every quantum resource effectively.

Quantum programs, typically expressed as quantum circuits, are designed
to require a certain number of qubits and gates, collectively
representing their resource demand. While algorithm developers aim to
reduce a program's resource usage from a domain's perspective, in the
NISQ-era, there is a heavy burden on the compilation layer to optimize
these circuits. Compilers like Qiskit~\cite{Qiskit}, Tket~\cite{sivarajah2020t}, and
BQSKit~\cite{osti_1785933} traditionally minimize gate or instruction count and
ignore the qubit count. Gate cancellation and removal involve
unitary synthesis~\cite{davis2020towards}, peephole-based template
matching~\cite{prasad2006data}, etc. All of
which cannot reduce circuit width (the number of physical qubits
required).

Mid-circuit measurement and reset (MMR) is a combination of primitive
operations that various quantum technology manufacturers have recently
integrated into their platforms, including
superconducting~\cite{corcoles2021exploiting},
trapped-ion~\cite{pino2021demonstration}, and
neutral-atom-based~\cite{graham2023mid} quantum hardware vendors. MMR's
original purpose was to implement quantum error correcting codes, but it
has also enabled new circuit optimization approaches in the NISQ-era.
These approaches optimize a circuit by reducing their required width in
a technique called circuit resizing, allowing users to execute larger
programs on cheaper, smaller quantum chips with potentially fewer gates.

Circuit resizing fundamentally works by reusing one physical qubit for
two program qubits, reducing the required qubit resources. To accomplish this, we schedule one qubit's operations on
the physical line and then measure and reset it, allowing the next qubit
to start its operations on this qubit. Not all circuits are resizable. Although several circuit resizing algorithms have been
proposed~\cite{brandhofer2023optimal, hua2023caqr, decross2022qubit,
sadeghi2022quantum}, they share common limitations: (1) They primarily
focus on circuit resizing, neglecting other optimization opportunities
arising during the process. (2) They only resize circuits that satisfy
specific gate dependence relationships, thereby overlooking
opportunities for resizing at the unitary level. This limitation significantly
restricts the range of programs that can benefit from this powerful
strategy.

This work introduces two novel resizing algorithms built on top of the Berkeley Quantum Synthesis Toolkit (BQSKit). The first combines
gate dependencies with traditional optimization strategies to reduce
both circuit width and depth with configurable parameters, leading to a
qubit count reduction of 45.3\% to 61.6\%. Second, we leverage
advancements in numerical instantiation to develop a resynthesis
algorithm that restructures non-resizable circuits into resizable ones.
This algorithm is resource-efficient and topology-aware, removing the
need for expensive mapping and increasing the potential for
optimization. We decrease the number of qubits by 20.7\% in previously
impossible-to-resize circuits while reducing gate counts by an average of
37.9\%. These resource optimizations make for an average improvement of 28.1\% improvement in
fidelity when executing on IBM's quantum machines.

\section{Background}
\label{sec:background}

\subsection{Numerical Synthesis and Instantiation}\label{sec:distance}
~\
Circuit synthesis converts a high-level description of a quantum program into an executable circuit. Since all quantum operations can be represented as unitary matrices, this process typically decomposes a large unitary into a set of small, native operations commonly consisting of one-qubit parameterized rotations in $U(2)$ and fixed two-qubit gates $U(4)$. Our focus is on unitary synthesis; throughout this paper, we will shorten this term to synthesis unless otherwise specified.

There are distinct exact and approximate methods for synthesis. Both
aim to synthesize circuits with as few gates as possible, but
approximate methods produce shorter circuits by allowing a small, configurable amount of error in the
calculation: $||U_T - U_S|| < \epsilon$. Here $U_T$ is the target
unitary matrix, and $U_S$ is the synthesized circuit's unitary. Every
algorithm will measure distance differently, but the leading algorithms
base their metric off the Hilbert-Schmidt inner product: $Tr(U_T^\dagger
U_S)$.

Practitioners commonly use the QSearch~\cite{davis2020towards}
approximate synthesis algorithm due to its topology-awareness,
efficacy, configurability, and composability. This synthesizer uses an
optimizer to tune gate parameters together with a search over circuit
structures to design efficient circuits automatically. This algorithm
refers to the parameter optimization as instantiation. Recently, QFactor,
a fast circuit instantiatier based on a tensor network formulation,
improved on this method by eliminating the need for explicit
parameterization~\cite{kukliansky2023qfactor}.

\subsection{Circuit Resizing}
\begin{figure}[h]
  \centering
  \includegraphics[scale=0.6]{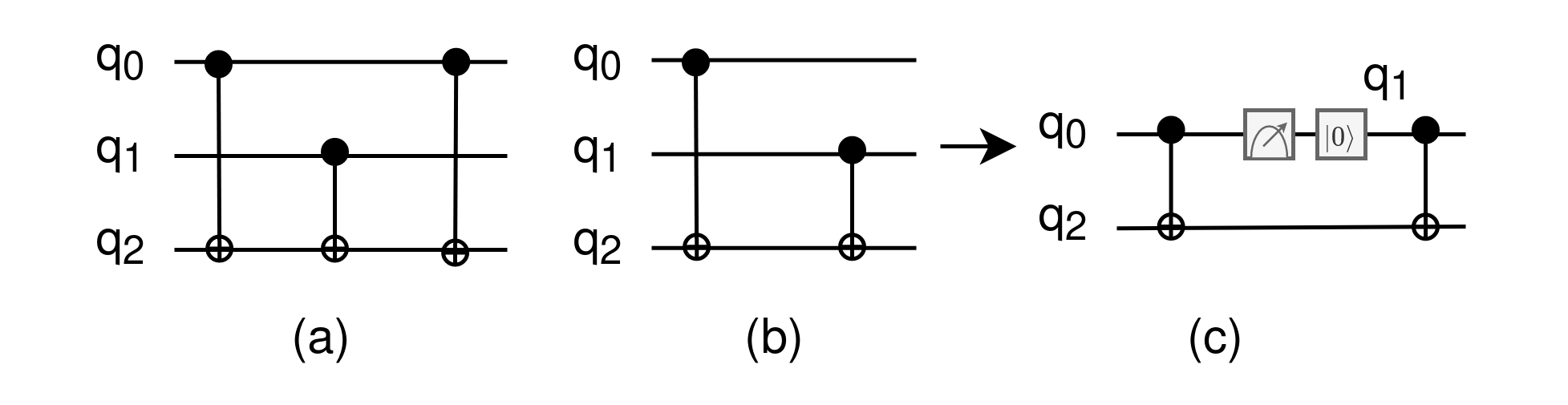}
  \caption{(a) This is a non-resizable circuit because there is no valid
      schedule of the gates that allows for one qubit to finish before
      another starts. (b) A resizable circuit because $q_0$ finishes
      before $q_1$ starts. (c) The previous circuit resized.
  }
  \label{fig:exp1}
\end{figure}

By effectively scheduling MMRs with qubit operations, we can decrease
the required number of qubits in a circuit. This optimization procedure
is called resizing. See Figure~\ref{fig:exp1} for an illustration of a
3-qubit program being resized to a 2-qubit one.

Circuits must satisfy specific
dependency properties for resizing to be possible. For two program
qubits to share one physical qubit, a schedule of gates must exist where
one of the qubit's instructions completes before the other's start. In
Figure~\ref{fig:exp1}a, there is no valid schedule of the gates that
allows for resizing, whereas in Figure~\ref{fig:exp1}b, program qubits,
$q_0$ and $q_1$ can share a physical qubit utilizing an MMR. 
Reducing the qubit counts of a quantum circuit offers several
benefits: (1) For larger quantum circuits, where the number of qubits
exceeds the capacity of the target quantum hardware, resizing enables circuit
execution. (2) Given that only a limited number of qubits possess high
fidelity for the near-term quantum hardware, reducing the circuit's
qubit count can exclude less reliable qubits, leading to improved
circuit fidelity. (3) For quantum hardware with nearest-neighbor
connectivity, such as IBM's superconducting quantum devices, fewer
qubits can reduce the number of SWAPs required during circuit
execution on the hardware. However, there are also challenges to
consider: (1) Resizing a quantum circuit using MMR may necessitate the
serial execution of more gates, potentially increasing the circuit depth
or duration. (2) The duration of MMR is longer than that of the other
gates. For example, on ibmq\_auckland, the average length of measurement is 3.8
times longer than that of a CNOT gate. Consequently, inserting MMRs may increase the overall circuit duration and introduce more
idle time. Balancing the advantages and limitations of circuit resizing
is crucial to maximizing the benefits derived from MMR.

\subsubsection{State-of-the-Art Resizing Algorithms}

Recently, several circuit resizing algorithms have been proposed, and
their effectiveness has been evaluated on
superconducting~\cite{brandhofer2023optimal,
sadeghi2022quantum,hua2023caqr} and trapped-ion quantum
devices~\cite{decross2022qubit}. DeCross et al. targeted the trapped-ion architecture with a SAT-solver approach for small circuits and a greedy heuristic for larger
programs, aiming for maximal qubit reuse~\cite{decross2022qubit}. In contrast, other methods primarily focused on superconducting quantum hardware. Sadeghi et al. presented 
a circuit resizing method to minimize output qubits but did not account for potential increases in circuit
depth~\cite{sadeghi2022quantum}. Hua et al. introduced the CaQR
compiler, which balanced the trade-off between the number of qubits
reused and the growth in circuit depth~\cite{hua2023caqr}. 
Brandhofer et al. incorporated
circuit resizing and chip connectivity conditions into an SMT
model, thereby achieving simultaneous circuit
resizing and mapping~\cite{brandhofer2023optimal}. However, all
these previous methods focused solely on analyzing gate dependencies
within the input circuit to identify resizing opportunities, without
considering other resource optimizations. Moreover, they did
not alter the circuit's structure through unitary synthesis
to modify gate dependencies and explore additional circuit resizing
possibilities.

\section{Gate-dependency Based Resizing}~\label{resizing}

Given a quantum circuit, we assess its resizability by gate dependency
analysis. If a circuit is resizable, we apply a search-based algorithm to minimize a configurable cost function described in this section. In the following section, we propose a method for resizing circuits that are not initially resizable.

The gate schedule in a circuit determines the gate-dependency rules for qubit resizing. We evaluate the qubit pair $(q_i, q_j)$ to check whether the completion of $q_j$ is independent of $q_i$, which would allow for the reuse of $q_i$ for $q_j$. To accomplish this, we traverse the circuit along $q_i$ until its final instruction and gather all qubits that interact with $q_i$. If a qubit $q_j$ is not seen, then the pair $(q_i, q_j)$ is recorded as a potential resizing opportunity. From this, we obtain a list of all resizable qubit pairs. This also gives us the  possible MMR insertion locations because in the resized circuit, we insert an MMR after $q_i$ to reuse it for $q_j$. If the list is not empty, the circuit is resizable based on the current gate dependencies. We can then apply our search-based algorithm to decide which qubits to reuse and where to apply MMRs; otherwise, we try the instantiation-based resizing check described in the following section.

Using a user-inputted cost function, we search over resizable pairs to determine the best-resized circuit. In this work, we consider two cost functions: one for maximal reuse, which aims to minimize the circuit qubit count as much as possible, and the second one for minimal depth, which aims to balance circuit width and depth optimization.

\begin{figure}[h]
    \centering
    \includegraphics[scale=0.50]{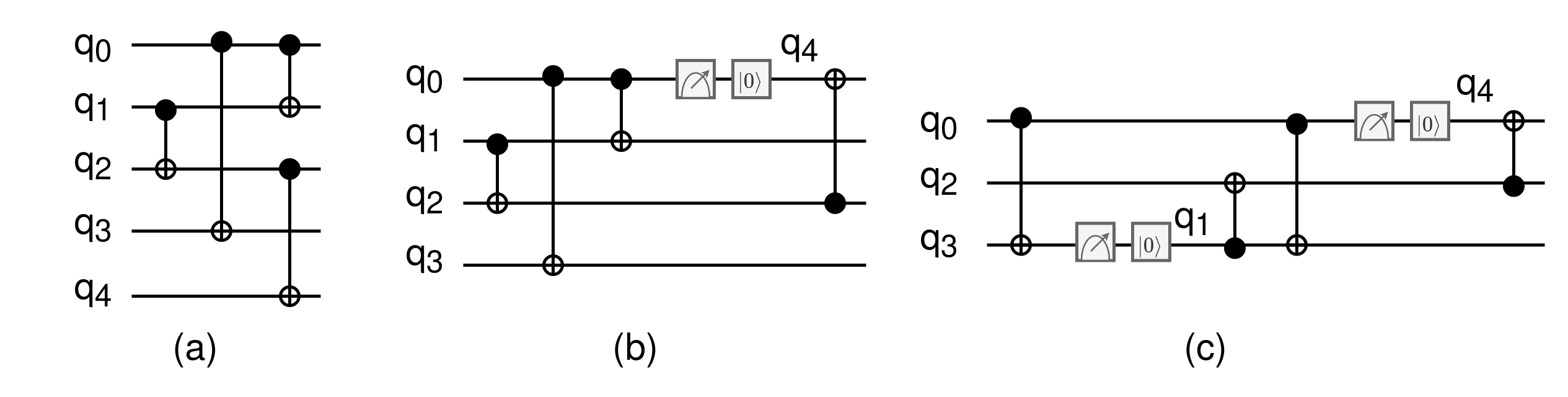}
    \caption{Given the 5-qubit input circuit, there are potential resizing opportunities via MMR insertions. Initially in (a), the potential resizing pairs are $(0, 4), (1, 4), (2, 0), (2, 3), (3, 1), (3, 2), (3, 4), (4, 0), (4, 3)$, where $(q_i, q_j)$ implies $q_i$ can be reused by $q_j$. We select (0, 4) to advance to (b), updating the remaining possible pairs to $(3, 1), (3, 2)$. Finally, we reuse $q_3$ for $q_1$ to obtain a 3-qubit circuit in (c), leaving no further resizing opportunities and terminating the algorithm.}
    \label{fig:resize_exp}
\end{figure}

Suppose there are only a few potential MMR insertion locations. In that case, the size of the search tree is small, allowing us to perform an exhaustive, brute-force search, ensuring we discover the best circuit. In the scenario with many resizing opportunities, we relax our search to a greedy-heuristic one. Here, we evaluate each resizing opportunity independently at each step, calculate the cost, and continue with the best-scoring circuit. One MMR is applied during each iteration, and the remaining potential resizing opportunities are updated to reflect the change. This procedure continues until no more resizing pairs can be found or the cost function cannot be reduced further. This heuristic resizing algorithm is efficient and ensures the best local solution at each MMR insertion step rather than guaranteeing a globally optimal solution. Figure~\ref{fig:resize_exp} depicts an example of resizing a quantum circuit.

We further optimize the circuit by synthesis once we obtain the optimally resized circuit according to our search-based approach. The resized circuit can be easily segmented into multiple parts separated by
MMRs. Within each partition, we re-synthesize the corresponding
sub-circuit using QSearch and perform gate deletion to reduce the number of gates. After replacing the subcircuits with their optimized outputs, we return the final resized circuit.

\section{Instantiation-based Resizing}~\label{resizable}

Instantiation, commonly used in unitary synthesis, enables one to separate a circuit's function -- the unitary it implements -- from its structure -- its decomposition into gates. We can vary the structure of a circuit to coerce it to something more amenable to resizing. This process facilitates finding non-intuitive circuit optimizations beyond simple commutativity or pattern-matching rules. In this section, we first describe how we use instantiation to check if the program that a circuit implements could ever be built in a resizable way. If we get a positive result, we then use a novel numerical-instantiation-based synthesis algorithm to resize the circuit for any native gates.

\subsection{Resizable Checking via Instantiation}

\begin{figure}
    \centering
    \includegraphics[scale=0.6]{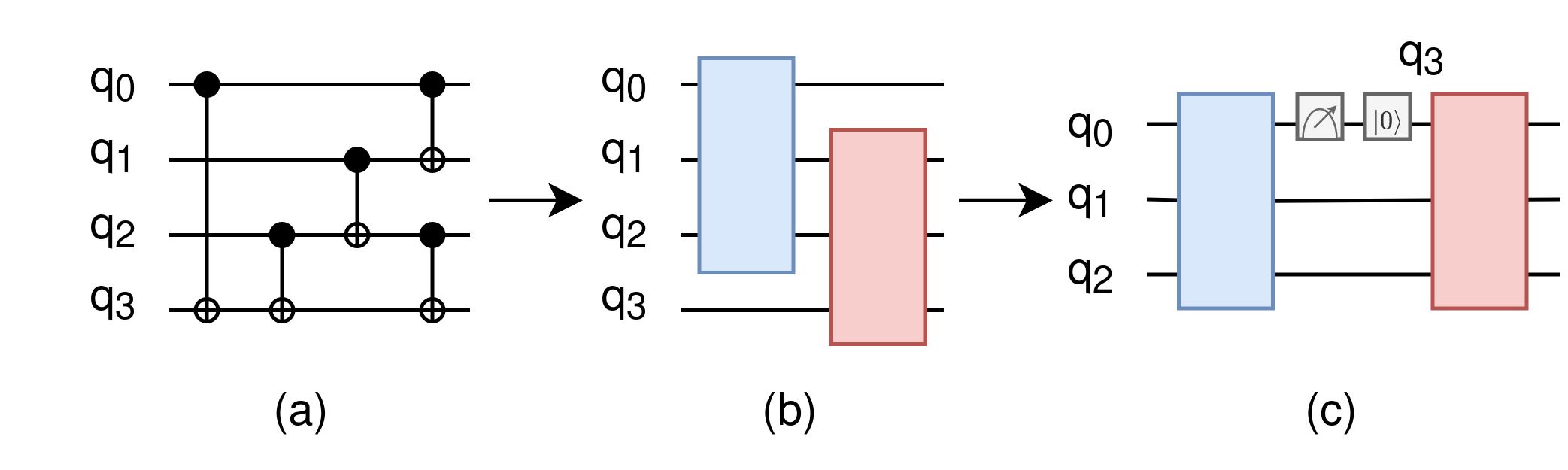}
    \caption{
The input circuit in (a) is not resizable via gate dependencies. Rather than work with the original circuit, in (b), we employ the instantiation-based resizing algorithm to check if $q_0$ can be reused for $q_3$ in some circuit that implements the original unitary. Here, each box represents a variable unitary matrix. If we successfully instantiate the template, we can resize the program as shown in (c).
    }
    \label{fig:exp-resize}
\end{figure}

Previously, we used gate dependencies to {\it find} resizing opportunities in a given circuit. This checking procedure requires a pre-built circuit and suffers from being downstream of a domain-specific circuit generator, which does not consider resizing. Here, we throw away the circuit structure entirely and {\it create} resizing opportunities by forming a fundamentally resizable, parameterized circuit. With the successful instantiation of this specific parameterized circuit to the original function, given as the program's unitary matrix, we know that this program is resizable at the unitary level. That is, there exists some circuit that implements the program's unitary that is resizable. Figure~\ref{fig:exp-resize} illustrates an example and the parameterized circuit style.

During resize-checking, our parameterized $n$-qubit circuits consist of two arbitrary unitary gates with $n-1$-qubits. We place the first block on all qubits except for $q_j$ and the second block afterward on all qubits except for $q_i$. Therefore, this template parameterizes all resizable circuits where $q_i$ can be reused for $q_j$. We then employ the QFactor tool to instantiate the circuit to the original unitary. We chose QFactor because it does not require explicit parameterization of variable unitary matrices. This feature makes it especially fast at solving these instantiation problems with rapid convergence. If the resulting instantiated circuit has a distance of less than some configurable epsilon to the input, we have successfully created a resizing pair for this algorithm. The distance is measured as explained in Section~\ref{sec:distance}. 

We evaluate each qubit pair of the input circuit, amounting to $n(n-1)$ parallelizable instantiation calls, and collect all successful outcomes. Before proceeding to the following synthesis step, we employ QFactor to downsize blocks in each successful circuit, leading to a higher-quality, accelerated synthesis step. We can reduce the size of parametrized blocks by removing an arbitrary qubit and reinstantiating like before. We continue until we have minimally sized blocks for each resizing pair. Finally, we select the pair with the smallest blocks to resynthesize.

\subsection{Circuit Resizing as Bottom-up Resynthesis}

After checking the resizability with instantiation, downsizing the blocks, and selecting a resizing pair $(q_i, q_j)$, we start our synthesis with an $n$-qubit circuit composed of two parameterized unitary blocks of at most $n-1$-qubits. These blocks ensure the reuse of $q_i$ for $q_j$. The next step is to synthesize these blocks into native gates provided by the user. We propose modifying the QSearch algorithm to decompose the two blocks into native gates in a topology-aware manner, removing the need for expensive mapping.

QSearch is a bottom-up synthesis strategy that utilizes $A^{*}$ to search over circuit structures and numerical instantiation to evaluate each structure. The algorithm incorporates an additional layer of configurable parameterized gates into the circuit with each deeper step in the search tree. This process terminates after discovering a program design successfully instantiating the target unitary. By only allowing valid instructions from a given topology, this process is made topology-aware, i.e., all gates in the final circuit occur between qubits connected in a target quantum chip. 

\begin{figure}
    \centering
    \includegraphics[scale=0.6]{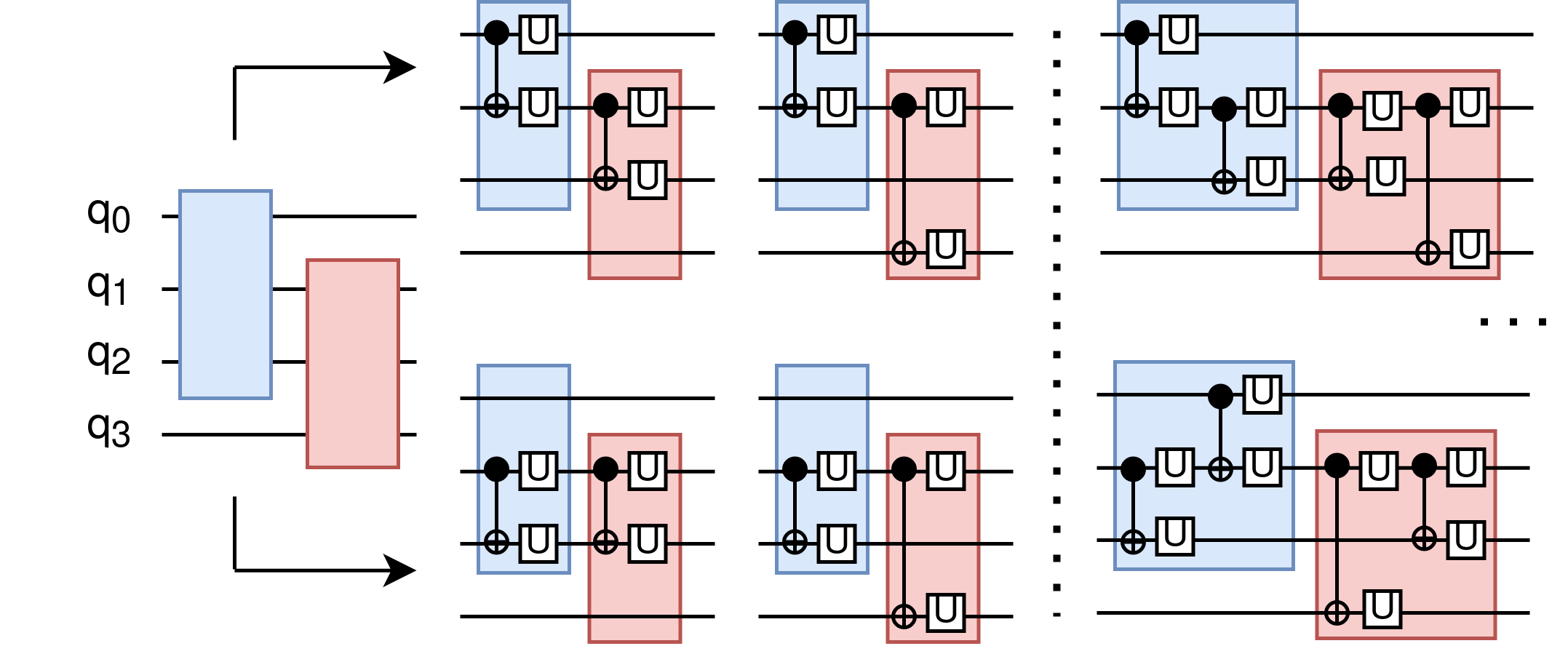}
    \caption{
This example demonstrates how we expand the input block-circuit from resize-checking into native gates. We target a chip with linear topology in the final circuit and the CNOT and U3 gate set. Every small box on one wire represents a U3 gate, a fully parameterized single-qubit rotation. With every expansion, we get four successors by adding one possible group of gates in every valid location.
    }
    \label{fig:synthesis}
\end{figure}

Our novel adaptation changes how QSearch builds its circuit structures to expand gates within the bounds of the input blocks. This modification maintains the resizability property of the circuit while allowing QSearch to modify parameters globally in the circuit. Naively, we could synthesize each block independently with QSearch off-the-shelf; however, this produces longer gate sequences due to a limited instantiation scope -- the area of variable parameters in a circuit. In other words, instantiation and synthesis tools will find higher-quality solutions when operating in a larger space, i.e., n-qubit circuit versus n-1-qubit block. Figure~\ref{fig:synthesis} demonstrates how we expand the blocks into native gates.

\begin{figure}
    \centering
    \includegraphics[scale=0.6]{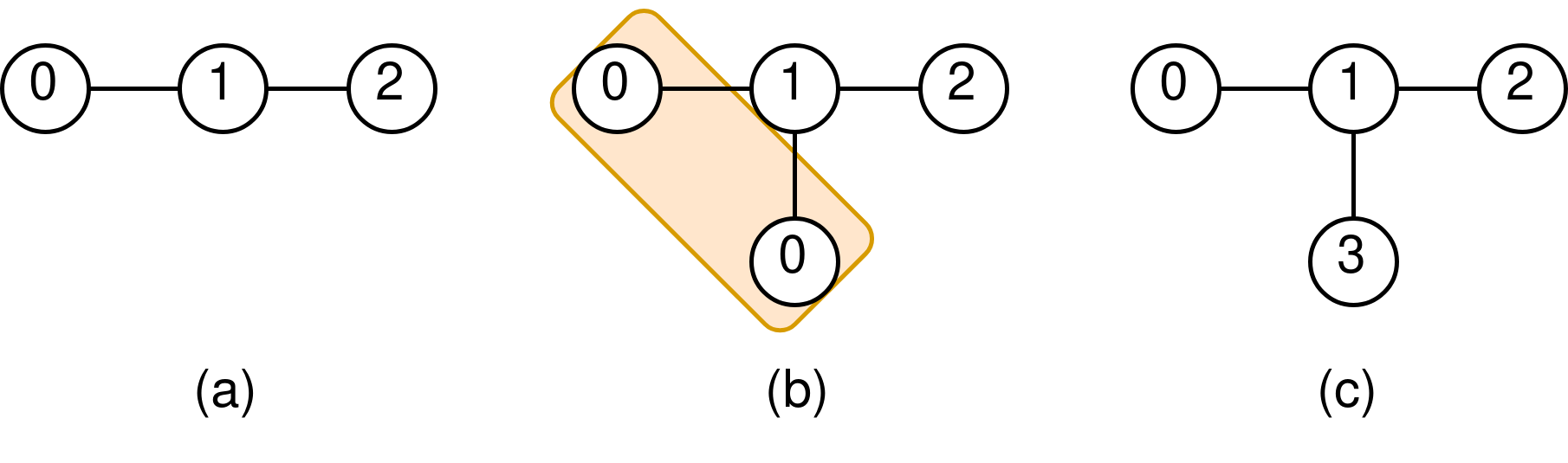}
    \caption{
This example showcases how we derive the connectivity restraints for the 4-qubit input circuit from the target 3-qubit chip connectivity. This process is necessary because we synthesize circuits before they are resized, and the resizing process changes the program's qubit connections. We start in (a) with the target coupling given by the quantum chip. In this example, we reused $q_0$ for $q_3$, so we fragment the zero node in (b). Finally, in (c), we relabel the fragmented node to the reused qubit, giving us the reverse-engineered topology. In the pre-resized circuit, gates between $q_1$ and $q_3$ will map to valid gates between $q_1$ and $q_0$.
    }
    \label{fig:enter-coupling}
\end{figure}

By default, QSearch is topology-aware, yet we must pay special attention to how we expand our circuit structures to maintain this property because our circuit will eventually be resized. Since we are synthesizing the circuit before resizing it, we must ensure that our allowed two-qubit gate interactions will reflect the reality after resizing. Let's assume we have a four-qubit problem, with $q_3$ being reused on $q_0$, and are targeting a chip with linear connectivity. If we allow linear interactions in our synthesis, then after resizing, we will replace the gates between ($q_2, q_3)$ with gates between $(q_2, q_0)$, which are illegal instructions. To combat this, we reverse-engineer a pre-resized topology that allows connections between qubits$(q_3, q_1)$ but not $(q_3, q_2)$. Now, after resizing, these gates will be valid linear interactions. Figure~\ref{fig:enter-coupling} displays how we reverse engineer the topology.

After synthesis, we finalize our result by performing gate deletion for further optimization. At this stage, the circuit is optimized and resizable. Since we selected the resizable pair $(q_i, q_j)$ before and maintained it throughout the synthesis process, inserting an MMR to $q_i$ and reusing it for $q_j$ completing the circuit resizing algorithm is straightforward.

\section{Experimental Setup}

We implemented both algorithms on top of version 1.1 of the Berkeley Quantum Synthesis Toolkit~\cite{osti_1785933}, a compilation framework using Python~3.11.4. We utilized BQSKit's implementation of the QSearch and QFactor techniques to accelerate the development of our algorithms. We used QSearch as-is during post-processing in our gate-dependency-based resizer, whereas we directly modified QSearch in our numerical-instantiation-based resizer. We called QFactor with default settings and used an epsilon of $10^{-10}$ for all instantiations calls throughout the evaluation. Our source code will be made available publically on GitHub.

The benchmarks collected from~\cite{quetschlich2023mqtbench,
zulehner2018efficient}, cover a variety of quantum algorithms, such as
Variational Quantum Eigensolver (VQE), Quantum Approximate Optimization Algorithm (QAOA), and Bernstein-Vazirani (BV), alongside applications in quantum arithmetic and error correction, among others. Note that QAOA circuits are generated based on two-regular graphs.

To evaluate the efficacy of our proposed resizing algorithms, we
consider the following metrics: the number of qubits required by the
circuits, the number of two-qubit gates, and the circuit
depth. The depth only considers two-qubit gates in the critical path to remove the impact of the variance of single-qubit gates due to the different basis gate sets. Furthermore, we employ two distinct metrics when assessing the fidelity of programs executed on quantum hardware. For circuits
that ideally produce a single correct output in the absence of noise, we use the Probability of Success Trial (PST), defined as the proportion of trials yielding the correct result out of the total conducted. Conversely, for circuits where the output is a probabilistic distribution, we utilize Hellinger fidelity to quantify the closeness of the experimentally obtained distribution on real quantum hardware to that predicted by ideal simulations.

We compare our algorithms against state-of-the-art compilation tools: BQSKit, Qiskit, and Tket. For each, we use the highest level of optimization, which is level 3 for qiskit, level 4 for BQSKit (which implements the PAM algorithm~\cite{liu2023tackling}), and level 2 for Tket. We targeted the two ibmq\_auckland and
ibm\_hanoi quantum IBM Q architectures for large circuits. When executing on real machines, we targeted the linear- and T-topologies for 3-5 qubit circuits.

\section{Results and Analysis}

\begin {table*}[t]
\begin{center}
\footnotesize
\begin{tabular}{|l|l|l|l|l|l|l|l|l|l|l|l|l|l|l|}
\hline  
\multicolumn{3}{|c|}{Benchmarks} & \multicolumn{2}{c|}{Tket} & \multicolumn{2}{c|}{Qiskit} & \multicolumn{2}{c|}{BQSKit} &
\multicolumn{3}{c|}{Maximal reuse} & \multicolumn{3}{c|}{Minimal depth}
\\
\hline
name & $n$ & CX & depth & CX & depth & CX & depth & CX & $n$ & depth & CX & $n$ & depth & CX \\
\hline
4mod & 5 & 10 & 12 & \bf{12} & 16 & 17 & 10 & \bf{12} & 3  & 12  & \bf{12} & 3 & 18 & 18\\
\hline
multiply-13 & 13 & 40 & 65 & 97 & 46 & 81 & 59 & 87 & 5 & 66  & \bf{73} & 7 & 73 & 91\\ 
\hline
system-9 & 12 & 148 & 256 & 338 & 252 & 338 & 171 & 239 & 5 & 223  & \bf{231} & 7 & 195 & 264\\
\hline
BV & 10 & 9 & 27 & 29 & 17 & 17 & 12 & 15 & 2 & 9  & \bf{9} & 4 & 13 & 18\\
\hline
QAOA-5 & 5 & 10 & 16 & 23 & 11 & 14 & 13 & \bf{13} & 3 & 13  & \bf{13} & 3 & 13 & \bf{13}\\
\hline
QAOA-10 & 10 & 20 & 28 & 30 & 21 & 32 & 23 & 33 & 3 & 20  & \bf{20} & 5 & 24 & 29\\
\hline
DJ & 10 & 9 & 27 & 29 & 17 & 22 & 12 & 15 & 2 & 9 & \bf{9} & 4 & 14 & 15\\
\hline
routing & 12 & 33 & 15 & \bf{33} & 15 & \bf{33} & 46 & 71 & 4 & 67  & 102 & 7 & 33 & 60\\
\hline
tsp & 9 & 40 & 16 & 40 & 16 & \bf{40} & 51 & 75 & 6 & 67  & 106 & 7 & 34 & 82\\
\hline %
\end{tabular}
\caption{Compiling circuits to IBM quantum hardware using gate-dependency-based resizing algorithm. $n$: qubit number. $CX$: total number of CX gates. depth: length of the critical path excluding single-qudit gates. 
}
\label{res:algo1}
\end{center}
\end{table*}

\begin{table}[h]
\begin{center}
\footnotesize
\begin{tabular}{|l|l|l|l|l|l|l|l|l|l|l|}
\hline  
\multicolumn{3}{|c|}{Benchmarks} & \multicolumn{2}{c|}{BQSKit\_L} &
\multicolumn{2}{c|}{Resize\_L} & \multicolumn{2}{c|}{BQSKit\_T} &
\multicolumn{2}{c|}{Resize\_T}\\
\hline
name & $n$ & CX & $n$ & CX & $n$ & CX & $n$ & CX & $n$ & CX\\
\hline
adder & 4 & 10 & 4 & 16 & 3 & 14 & 4 & 17 & 3 & 14\\
\hline
vqe & 5 & 25 &  5 & 55 & 4 & 35 & 5 & 52 & 4 & 28\\ 
\hline
qec & 5 & 11 &  5 & 25 & 4 & 11 & 5 & 27 & 4 & 12\\
\hline
decod & 5 & 27 & 5 & 36 & 4 & 27 & 5 & 42 & 4 & 27\\
\hline
alu & 5 & 32 & 5 & 41 & 4 & 40 & 5 & 45 & 4 & 25\\
\hline
mod5 & 5 & 22 & 5 & 45 & 4 & 19 & 5 & 30 & 4 & 16\\
\hline %
\end{tabular}
\caption{Compiling circuits to linear and T topology using numerical-instantiation-based resizing algorithm. All the circuits are not resizable by any other tool.
}
\label{res:algo2}
\end{center}
\end{table}

\begin{figure}
    \centering
    \includegraphics[scale=0.28]{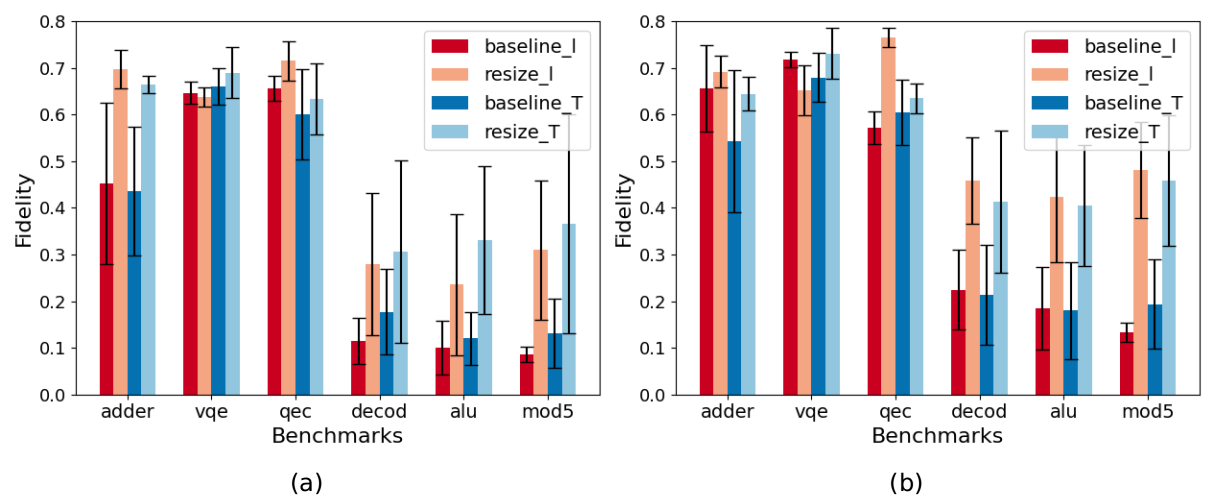}
    \caption{Circuit fidelity results from executions on (a) ibmq\_auckland and (b) ibmq\_hanoi.
    }
    \label{fig:hardware-res}
\end{figure}

Table~\ref{res:algo1} details the comparison between our gate-dependency-based circuit resizing algorithm and state-of-the-art tools. When the cost function is optimized for maximal qubit reuse, there is a substantial 61.6\% reduction in qubit count. In all benchmarks, except for routing and tsp, reducing the qubit count leads our resizer to decrease CNOT gate count by 11.4\% compared with the best of the other compilers. This does increase the circuit depth by 22.4\%. However, when the cost
function prioritizes minimizing depth, the
resulting depth is now only increased by 5\%. In
this scenario, the qubit count is lowered by 45.3\%. The minimal depth cost increases the CNOT gate count by 2.6\% compared to maximal reuse. These experimental results
effectively highlight the trade-off between circuit depth and the number
of qubits required. The slight increase in CNOT gates for our gate-dependency-based resizer, compared with BQSKit stems from the differences in the mapping algorithms, SABRE~\cite{li2019tackling} and PAM respectively. We anticipate adapting PAM to our resizer to potentially further reduce the number of CNOT gates.

We also compare our gate-dependency-based circuit resizer with CaQR~\cite{hua2023caqr}, another circuit resizing method also pursuing maximal reuse and minimal circuit depth metrics. Due to the inaccessibility of their source code, executable, and benchmark dataset, our comparison is confined to the four benchmarks (4mod, multiply-13, system-9, BV) with extractable results from their paper. When maximizing qubit reuse, our method achieves an additional reduction of 28.6\% in qubit count and 21.4\% in the total number of CNOT gates in contrast to CaQR. In the scenario prioritizing minimal circuit depth, we observe a decrease of 25\% in qubit count and 8.2\% in CNOT gates. Our approach, however, increases depth by 47\% and 43\% respectively when counting the overal circuit depth including both single and two-qubit gates. This is due to our substantial use of off-the-shelf QSearch synthesis, which is primed to reduce two-qubit gate count without consideration of single-qubit depth. CaQR calculates circuit depth to include both single- and two-qubit gate critical path. If we tune QSearch to minimize depth, the algorithm will produce shallower circuits at potentially the cost of more gates. The current trade-offs are beneficial since two-qubit gate error dominates in NISQ devices, but we can tune as this changes.

Table~\ref{res:algo2} presents the results of compiling circuits to linear and T topology using our numerical-instantiation-based resizing algorithm. This approach enabled the reuse of one qubit per tested benchmark, yielding an average reduction in circuit size by 20.7\%. Compared with the BQSKit of linear and T topology, the number of two-qubit gates decreased by 33\% and 42.7\%. To facilitate fair execution on IBM quantum hardware, we carefully chose three placements that align with the linear and T topologies while avoiding qubits and connections with high error rates. We then mapped the benchmarks to these placements. As reported by~\cite{brandhofer2023optimal}, the number of reset repetitions affects fidelity. To assess this, we vary the resets from one to three. The experimental results show that a single reset yields the highest fidelity, surpassing two and three resets by 17.1\% and 26.9\%, respectively. Therefore, in Figure~\ref{fig:hardware-res}, we report only the single reset results.

Resizing enhanced the fidelity of all tested circuits and topologies except for the linearly compiled VQE circuit. The fidelity of the VQE circuit was high before resizing, making it difficult to improve. Overall, for ibmq\_auckland, the circuit fidelity is enhanced by 28.5\% and 28.9\% for linear and T topologies respectively. Similarly, for ibmq\_hanoi, improvements are noted at 28.4\% for linear topology and 26.6\% for T topology.
\section{Discussion and Conclusion}
\label{sec:conclusion}

Mid-circuit measurement and reset provide circuit optimization opportunities by circuit resizing, which lowers the required number of qubits, reducing required gates and enhancing fidelity. This paper introduced two resizing algorithms: one that leverages search and gate dependencies and another that uses numerical instantiation to find non-intuitive restructures. We conclude with two brief discussions, one on runtimes and scalability and another on the search-based cost function.

For the gate-dependency-based resizer, we use brute-force search for circuits with fewer than 7 initial resizable pairs; for circuits with more resizing opportunities, we switch to the heuristic method. This empirically-informed decision guarantees that runtimes for the gate-dependency-based algorithm are within seconds and that it is scalable to large system sizes. The gate-dependency-based approach, however, requires circuits to have resizable opportunities before application, whereas the numerical-instantiation-based algorithm does not have the same restriction. Our resynthesis prototype has successfully demonstrated its effectiveness on circuits up to five qubits, showcasing the strength of instantiation in restructuring circuits for better optimization potential. Instantiation does incur an exponential scaling due to the optimization of larger systems; QFactor was shown to scale to 12 qubits directly with GPUs. Prior works have overcome this challenge with divide-and-conquer strategies using vertical circuit partitioning scaling instantiation-based methods to thousands of qubits. In a production compiler pipeline, we envision a scalable approach using a similar paradigm but leave this to future work.

Mid-circuit measurements and resets come with a substantial physical cost. On some hardware platforms, measurements are longer in duration and more error-prone than unitary gates. For example, in superconducting circuits, typical gate times are $\sim$ 10 ns for single-qubit gates and $\sim$ 100 ns for two-qubit gates, whereas measurements typically take anywhere from $\sim$ 500 ns to a few $\mu$s. Therefore, naive application of mid-circuit measurements can drastically increase the total execution time of a quantum circuit. Moreover, mid-circuit measurements can incur errors on ``active'' spectator qubits during their execution \cite{govia2022randomized}, increasing their total cost. In this work, we demonstrated an improvement in fidelity using our simple maximal reuse or minimal depth cost models. Looking forward, with a more meticulous application of our algorithm, a user may tweak the cost function to include these additional physical constraints, such as circuit duration or spectator errors.

\acknowledgments

This work was supported by the U.S. Department of Energy, Office of Science, Office of Advanced Scientific Computing Research through the Accelerated Research in Quantum Computing Program. This research used resources of the Oak Ridge Leadership Computing Facility, which is a DOE Office of Science User Facility supported under Contract No.~DE-AC05-00OR22725. A.H.~acknowledges financial support from the U.S.~Department of Energy, Office of Science, Office of Advanced Scientific Computing Research Quantum Testbed Program under Contract No.~DE-AC02-05CH11231.
The authors also acknowledge the use of IBM Quantum services. The views expressed are those
of the authors and do not reflect the official policy or position of IBM or the IBM Quantum team.
\bibliographystyle{ACM-Reference-Format}
\bibliography{sample}

\end{document}